# The structural and magnetic properties of $(In_{1-x}Fe_x)_2O_3$ (0.0 ≤ x ≤ 0.25) system : prepared by gel combustion method

O. D. Jayakumar, I. K. Gopalakrishnan[a] and S.K. Kulshreshtha

Chemistry Division, Bhabha Atomic Research Centre, Mumbai-400085, INDIA.

Amita Gupta, and K. V. Rao

Dept. of Materials Science-Tmfy-MSE, Royal Institute of Technology,

Stockholm SE 10044, Sweden.

D.V. Louzguine-Luzgin, and A. Inoue

Institute for Materials Research, Tohoku University, Sendai 980 Japan.

P.-A. Glans and J.-H Guo

Advanced Light Source, Lawrence Berkeley National Laboratory, Berkeley, CA 94720

K. Samanta, M.K. Singh, and R.S. Katiyar

Department of Physics, University of Puerto Rico, San Juan, Puerto Rico 00931-3343

**Abstract:**

$(In_{1-x}Fe_x)_2O_3$ polycrystalline samples with x = (0.0, 0.05, 0.10, 0.15, 0.20 and 0.25) have been synthesized by a gel combustion method. Reitveld refinement analysis of X ray diffraction data indicated the formation of single phase cubic bixbyite structure without any parasitic phases. This observation is further confirmed by high resolution transmission electron microscopy (HRTEM) imaging, and indexing of the selected-area electron diffraction (SAED) patterns, X-ray Absorption Spectroscopy (XAS) and Raman Spectroscopy. *DC* Magnetization studies as a function of temperature and field indicate that they are ferromagnetic with Curie temperature ($T_C$) well above room temperature.

[a] Corresponding author Email: ikgopal@barc.gov.in

The theoretical prediction of room-temperature ferromagnetism (RTF) in Mn doped p-type ZnO and GaN by Dietl[1] and co-workers triggered a worldwide research on dilute magnetic semiconductor (DMS) materials.[2-10] For practical spintronic applications DMS with Curie temperature, $T_c$, above RT are required. Many promising oxides like $TiO_2$, ZnO and $SnO_2$ and III-V semiconductor GaN doped with transition metal ions have been reported to exhibit ferromagnetic behavior near or above room temperature,[3-18] but the origin of ferromagnetism in these systems is poorly understood. Oxide systems based on host semiconductor with high solubility of magnetic ions are highly desirable to form thermodynamically stable magnetic semiconductors. Yoo et al.[19], in search of oxide based spintronic materials, have discovered by combinatorial methodology a ferromagnetic semiconductor system in $(In_{1-x}Fe_x)_2O_{3-y}$ (x up to 0.20), co-doped with Cu. $In_2O_3$ is a wide band gap (3.7eV) transparent n-type semiconductor. They have reported RTF in bulk samples of Cu co-doped $(In_{1-x}Fe_x)_2O_{3-y}$ for $x \geq 0.10$, prepared by high temperature solid state reaction of $In_2O_3$, $Fe_2O_3$ and CuO. Later, He et al.[20] reported preparation of thin films of Cu co-doped $(In_{1-x}Fe_x)_2O_{3-y}$. It was also shown that in bulk and thin films, RTF disappears if it is not co-doped with Cu or when subjected to annealing in oxygen atmosphere. Recently, Philip et al.[21] have reported different magnetic phases for Cr doped $In_2O_3$ films, obtained by varying the carrier density with the defect concentration. Very recently, Gupta et al.[22] showed that V and Cr doped $In_2O_3$ thin films and bulk, prepared by both sol gel and solid state reaction are ferromagnetic with $T_c$ well above room temperature. In this paper, we report the synthesis of polycrystalline $(In_{1-x}Fe_x)_2O_3$ samples, showing RTF without any co-doping, prepared by gel combustion method.

Fe doped $In_2O_3$ was made by combustion method using glycine as the fuel and appropriate amounts of nitrates of Fe and In. Phase purity and the structure of the fine powder obtained were analyzed by powder X-ray diffraction using CuKα radiation

HRTEM investigations were carried out using a JEM 2010 (JEOL) microscope operating at 200 kV equipped with an energy dispersive X-ray spectrometer (EDX) of 0.1 keV resolution. The X-ray absorption spectroscopy (XAS) experiment was performed at beamline 7.0.1 of Advanced Light Source. XAS spectra were obtained by measuring the total electron yield (TEY) from the sample as a function of the incoming photon energy. All spectra were normalized to the photocurrent from a clean gold mesh introduced into the beam. The resolution was set to 0.2 eV for both the O *K*-edge and Fe *L*-edge X-ray absorption measurements. Raman spectra of $In_2O_3$ and 10% Fe doped $In_2O_3$ were obtained using T64000 Raman microprobe system from Horiba, Inc. with an Ar+ ion laser source of 514.5 nm wavelength. *DC* magnetization measurements as a function of field and temperature were carried out using an E.G.&G P.A.R vibrating sample magnetometer (model 4500) and Quantum Design SQUID magnetometer (model MPMS).

Rietveld profile refinement analysis of XRD data of $(In_{1-x}Fe_x)_2O_3$ (x = 0.0, 0.05, 0.10, 0.15, 0.20 and 0.25) samples showed that they are single phase with cubic bixbyite structure (space group Ia3) and lattice parameter *a* decreased linearly with increasing Fe concentration up to *x* = 0.2, obeying Vegard's law, confirming the incorporation of Fe into $In_2O_3$ lattice (Graph 1 and 2 supplementary material).[19][5] However, Yoo et al.[19] reported the appearance of an impurity phase of $Fe_3O_4$ in their samples prepared by solid state reaction of oxide constituents. On the other hand, Gurlo et al.[23] have observed hexagonal $In_2O_3$ when the 'Fe' concentration exceeded 10 at % in Fe doped $In_2O_3$, prepared by sol-gel method. This may be due to the method of synthesis used by them. However, in the present method of preparation used by us, up to 25 % of Fe, we did not observe any impurity phase of $Fe_3O_4$ and it retained its cubic bixbyite structure. Nevertheless, the Fe concentration '*x*' vs. lattice parameter plot deviated from linearity for x > 0.20 indicating the solubility limit of Fe in $In_2O_3$ lattice as ≈ 20 %.[23] Energy Dispersive X-ray (EDX) analysis carried out on $(In_{1-x}Fe_x)_2O_3$ samples showed that they are homogeneous and the cation concentrations found by EDX are in good agreement with the nominal cation concentrations within the experimental error.

Detailed structural characterization of 10 at % Fe doped $In_2O_3$ sample was performed by HRTEM imaging in order to investigate the possibility of secondary phase inclusions. It can be seen from the HRTEM image depicted in Figure 1a that there are no detectable traces of secondary phases or precipitates of any ferromagnetic oxides of Fe. The stripes in the HRTEM image correspond to the (211) family planes of the body-centered cubic bixbyite structure. The SAED pattern obtained from this grain is shown for illustration. Indexing of the SAED patterns matched with cI80 $In_2O_3$ lattice. Thus, both HRTEM image and SAED pattern show the grains are free of secondary phase intergrowths and nanoclusters.

Raman spectra of Fe 10 at % doped $In_2O_3$ and $In_2O_3$ are depicted in figure 1b. All of the prominent peaks of $In_2O_3$ are also observed in the Fe doped $In_2O_3$ spectra. These peaks are, however, slightly shifted towards the lower frequency and had considerable increase in their half widths. A comparison of the two spectra (doped and undoped) clearly indicates that Fe is indeed occupying interstitial/substitutional sites in the $In_2O_3$ lattice. The downward shifts in the peak values and the broadening of the modes can be interpreted as due to the lattice disorder at the nanocrystalline level.

Figure 2 (inset) shows the Fe L-edge XAS spectra of $Fe_2O_3$ and Fe doped $In_2O_3$. Fe $L_{2,3}$-absorption features are allocated in between 706 eV and 725 eV, with a spin-orbit splitting of approximately 13 eV. Both XAS spectra are alike, and the same splitting of about 1.5 eV for the first and second absorption peaks at $L_3$-edge indicated that Fe is in the 3+ oxidation state in $Fe:In_2O_3$ when Fe replaces In site. Figure 2 shows the O K-edge XAS spectra of $Fe_2O_3$, $In_2O_3$, and Fe doped $In_2O_3$, the difference spectrum of $In_2O_3$ and Fe doped $In_2O_3$ is also included. The oxygen XAS spectra, to a first-order approximation, reflect the unoccupied oxygen-p projected density of states. The well separated pre-edge in the XAS spectrum of $Fe_2O_3$ is attributed to oxygen *2p* weight of states that have predominantly transition-metal *3d* character [24], i.e. Fe *3d* - O *2p* mixing. The *3d* states is split into two bands, which are related to $t_{2g}$ and $e_g$ symmetry, although such division is not strictly valid as the crystal field is slightly distorted from octahedral symmetry. The overall O K-edge XAS spectral profile of Fe doped $In_2O_3$ is similar to that of $In_2O_3$, while conduction band (CB) edge is extended of 1.5 eV into the band gap. Thus, there are doped holes in the bottom of conduction band induced by Fe doping. For a detailed comparison, the difference spectrum was obtained by subtracting 90% of contribution of $In_2O_3$ as Fe concentration is 10% in the $Fe:In_2O_3$. The splitting between the first two peaks in the difference XAS spectrum is less than 1 eV in comparison with 1.5 eV of $Fe_2O_3$, which suggests that Fe atoms are located in the crystal structure of $In_2O_3$. Thus XAS data shows that Fe is in 3+ charge state and is incorporated into $In_2O_3$ crystal lattice.

Figure 3(a) depicts the room temperature *DC* magnetization measurements of $(In_{1-x}Fe_x)_2O_3$ samples as a function of field. It can be seen from the figure that all samples showed symmetric hysteretic loops typical of ferromagnetic materials. The saturation magnetization value *Ms* is found to increase with Fe concentration up to 20%. The compound with x = 0.25 showed slight decrease in *Ms* value. This shows that the maximum solubility limit of Fe in $In_2O_3$ matrix is 20 % of Fe in agreement with Yoo et al. [19]. Yoo et al. observed ferromagnetism at room temperature (RT) only for Cu co-doped $(In_{1-x}Fe_x)_2O_3$ with Fe concentration above 10%. They also found an initial decrease in *Ms* value with increasing Fe concentration up to 7%, which they attributed to antiferromagnetic $Fe^{3+}$-O-$Fe^{3+}$ pairing as isolated $Fe_2O_3$ molecule dissolved in the $In_2O_3$ lattice. In the present study, we could observe ferromagnetism for Fe concentration right from 5 to 25 % without any co-doping. It is worth mentioning that we have not treated the sample in oxygen deficient atmospheres as has been done by Yoo et al.[19]

Figure 3(b) shows the temperature dependence of the magnetization for the representative sample 10 % Fe : $In_2O_3$ in a field of 1 Tesla measured during the warming run after cooling the sample in the same field from above room temperature. The steep increase with decreasing temperatures below around 50K in the magnetization is characteristic of all DMS materials, and is probably related to the defects structure and possible fraction of Fe atoms which are not participating in the long range ferromagnetic order. The coercive field ($Hc$) is much larger (ranging from 200 to 300 Oe) in our samples compared to $Hc$ values (40-50 Oe) obtained by Yoo et al.[20] in their Cu co-doped $(In_{1-x}Fe_x)_2O_3$ samples prepared by solid state reaction at high temperature. Notice that the temperature dependence of the magnetization above 70 K can be fitted to the $T^{3/2}$ Bloch law for spin waves all the way to room temperature. Without knowing the exact concentration of the Fe atoms participating in the ferromagnetic order it is not meaningful to calculate the effective moment per Fe atom from the $M_s(0)$ value obtained from the remarkable fit over such a wide range. The reservation holds good in interpreting the spin wave stiffness value that is inversely proportional to the coefficient B of the $T^{3/2}$ term. However the order of magnitude, $10^{-5}$ $K^{-3/2}$, obtained for B is typical for disordered or amorphous systems, indicating that it is easier to propagate spin waves in these DMS material. It also indicates clearly that there is no contribution from any other possible crystalline magnetic phases like for example $Fe_3O_4/Fe_2O_3$. The coefficient B is one order of magnitude smaller in the case of crystalline materials arising from the higher values for the spin wave stiffness. This inference is consistent with HRTEM, XAS and Raman spectroscopy results, which indicate that the observed RTF in In-Fe-O system cannot be attributed to any possible presence of crystalline $Fe_3O_4$. It is important to point out that the weak temperature dependence of the magnetization extending over a wide range up to room temperature also indicates that the actual $T_C$ for the sample is far above room temperature.

In conclusion, polycrystalline samples of $(In_{1-x}Fe_x)_2O_3$ (x = 0.0, 0.05, 0.10, 0.15, 0.20 and 0.25) have been synthesized by a gel combustion method. Powder X-ray diffraction and HRTEM studies showed single-phase compound up to x = 0.20. XAS and Raman spectra of Fe doped and undoped $In_2O_3$ also indicate the incorporation of Fe into $In_2O_3$ lattice. Magnetization studies as a function of field and temperature showed RTF in these samples. The observed RTF may be attributed to the generation of defects during the synthesis of Fe doped $In_2O_3$.

## Acknowledgements

The work at the Advanced Light Source (LBNL) is supported by the Office of Basic energy Science of the US Department of Energy under Contract No. DE-AC02-05CH11231. Research in Sweden has been funded by the Agency VINNOVA.

**Figure Captions**

Fig.1 (Color online) (a) HRTEM image and indexing of the selected-area electron diffraction pattern of $(In_{0.90}Fe_{0.10})_2O_3$ (inset).

(b) Raman spectra of $In_2O_3$ (upper curve) and $(In_{0.90}Fe_{0.10})_2O_3$ (lower curve).

Fig.2. (Color online) O K-edge XAS spectra of $Fe_2O_3$, $In_2O_3$, and Fe doped $In_2O_3$, the difference spectrum of $In_2O_3$ and Fe doped $In_2O_3$ is also included. Inset shows Fe L-edge XAS spectra of $Fe_2O_3$ and Fe doped $In_2O_3$.

Fig.3 (a) (Color online) *M vs H* curves at RT for $(In_{1-x}Fe_x)_2O_3$ (x = 0.0, 0.05, 0.10, 0.15, 0.20 and 0.25). (Inset shows data near origin at an expanded scale).

(b) Temperature dependence of the Magnetization for $(In_{0.90}Fe_{0.10})_2O_3$ field cooled at 1Tesla and measured in the same field during the warming run. Data above 70K fits well to the Bloch's law for spin waves with a $T^{3/2}$ temperature dependence.

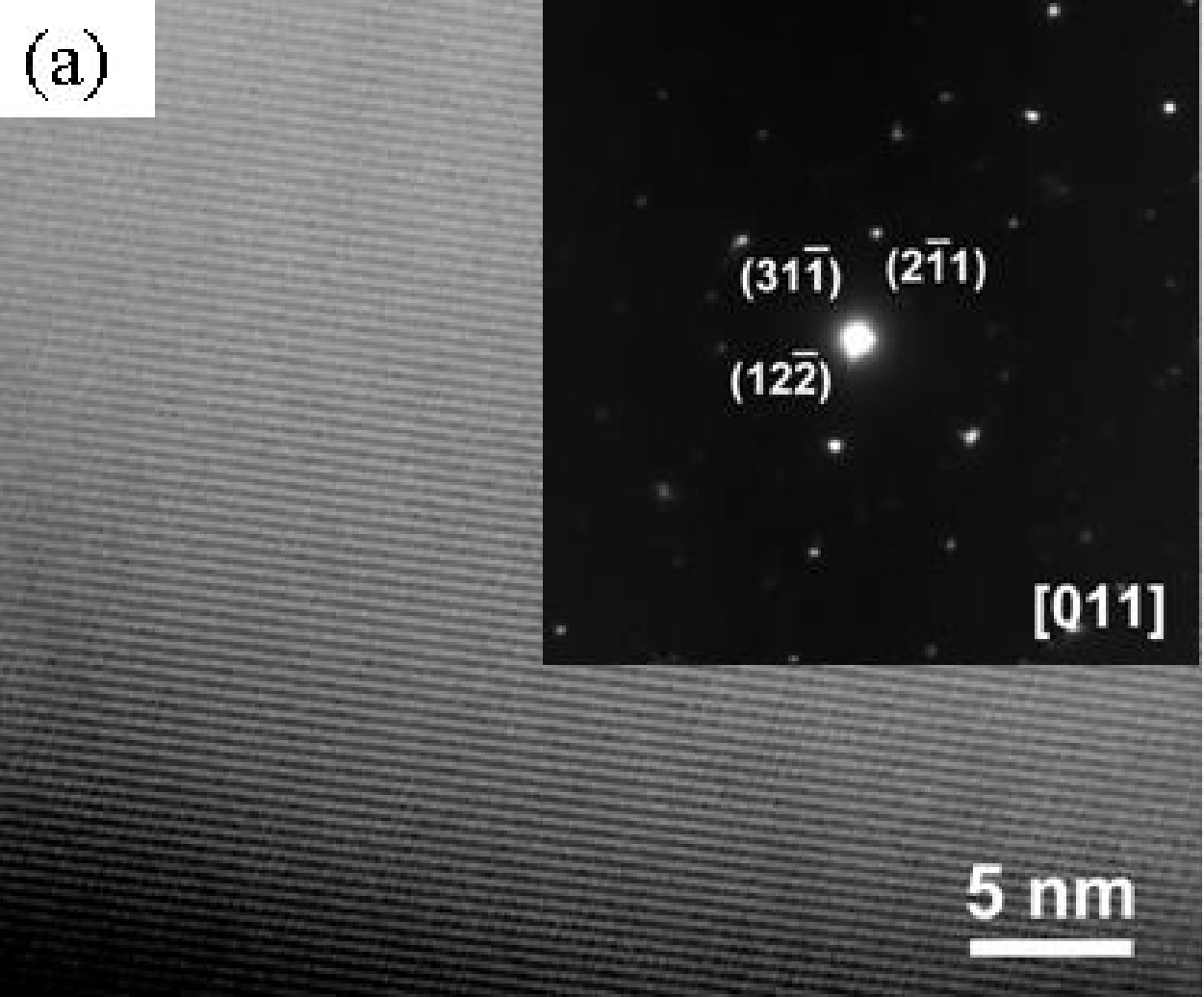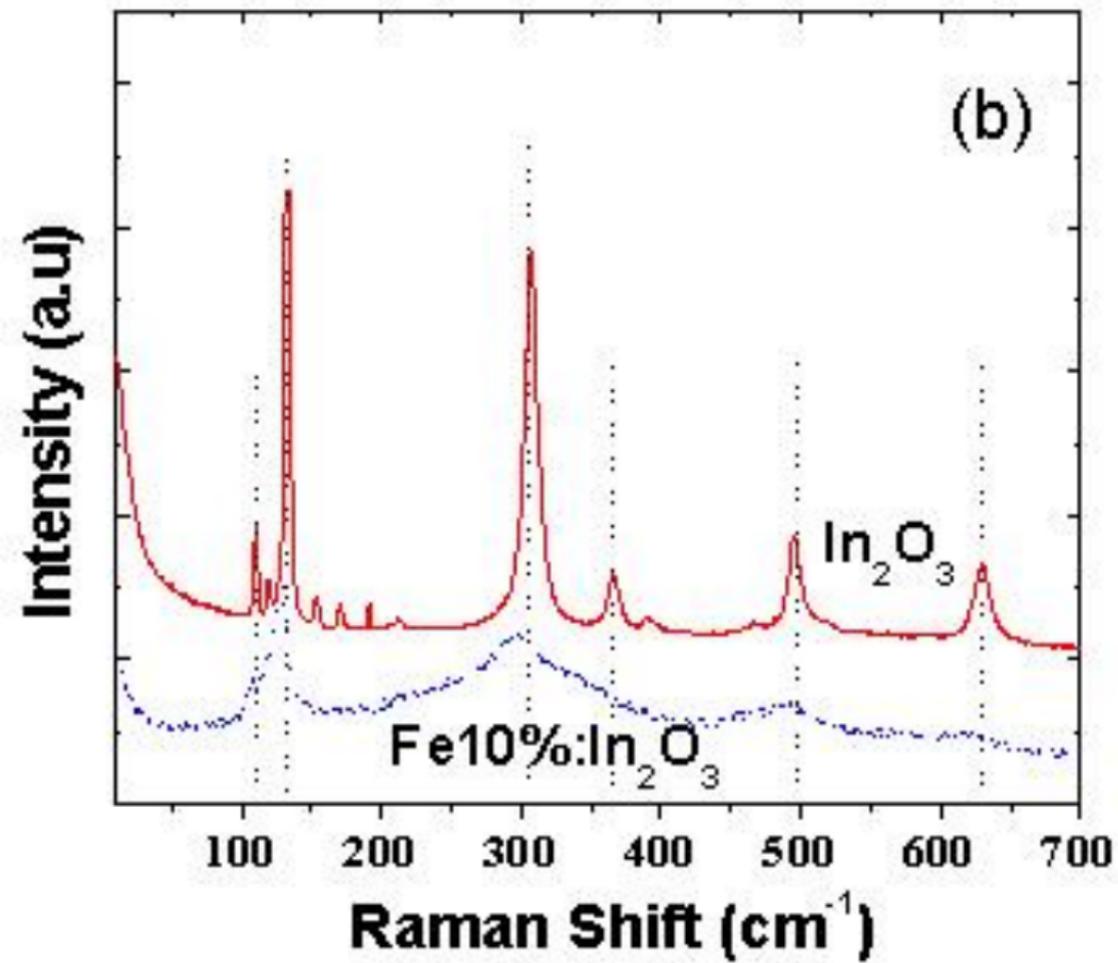

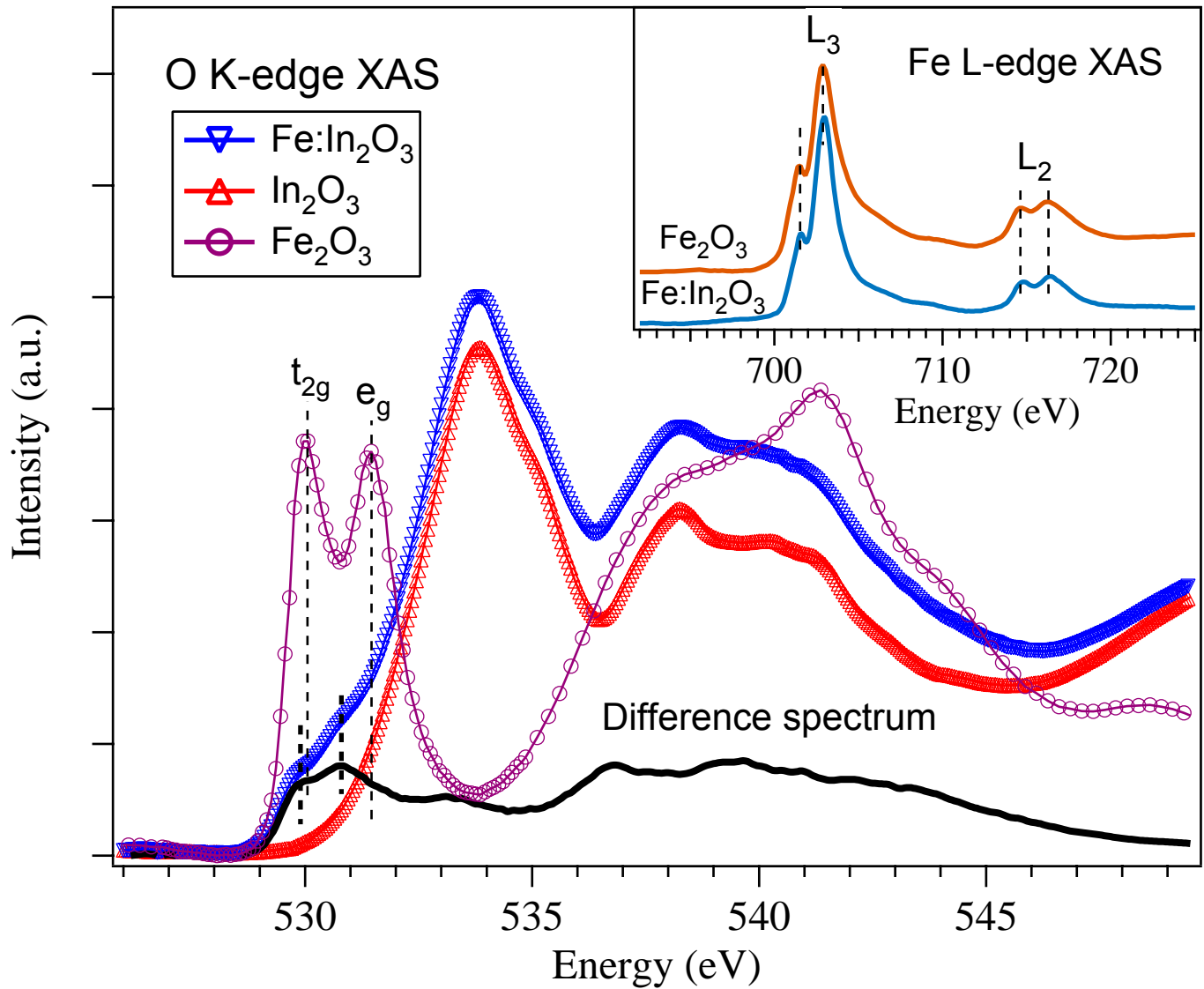

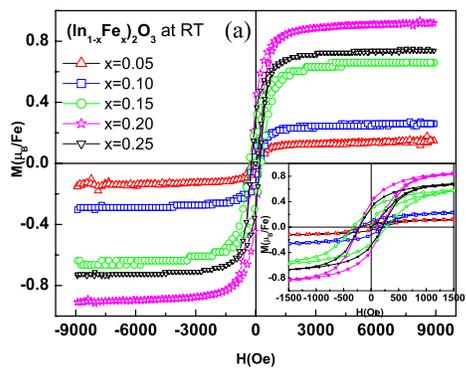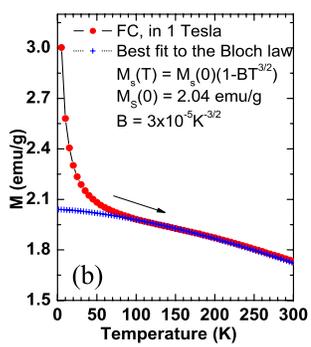